\documentclass[aps,prl,floatfix,twocolumn, showpacs]{revtex4}

\usepackage{amsmath}
\usepackage{graphicx}
\usepackage{epstopdf}
\usepackage{psfrag}
\usepackage[usenames]{color}
\usepackage{hyperref}
\usepackage{bm}
\usepackage{bbm}
\usepackage{soul}

\newcommand{\ketbra}[2]{\mbox{$|#1\rangle\langle #2|$}}

\newcommand{\ket}[1]{\vert#1\rangle}
\newcommand{\bra}[1]{\langle#1\vert}

\begin{document}

\title{Experimental feedback control of quantum systems using weak measurements}

\author{ G. G. Gillett$^{1\dagger}$, R. B. Dalton$^{1\dagger}$, B. P. Lanyon$^1$, M. P. Almeida$^1$, M. Barbieri$^{1,5}$, G. J. Pryde$^2$, J. L. O'Brien$^3$, K. J. Resch$^4$, S. D. Bartlett$^6$ and A. G. White$^1$}

\affiliation{$^1$ Department of Physics and Centre for Quantum Computer Technology, the University of Queensland, Brisbane 4072, Australia}
\affiliation{$^2$ Centre for Quantum Dynamics and Centre for Quantum Computer Technology,  Griffith University, Brisbane 4111, Australia}
\affiliation{$^3$ Centre for Quantum Photonics, H. H. Wills Physics Laboratory and Department of Electrical and Electronic Engineering, University of Bristol, Bristol BS8 1UB, United Kingdom.}
\affiliation{$^4$ Institute for Quantum Computing and Department of Physics and Astronomy, University of Waterloo, Waterloo, Ontario, N2L 3G1 Canada}
\affiliation{$^5$ Laboratoire Charles Fabry, Institut d'Optique, Palaiseau 91127, France.}
\affiliation{$^6$ School of Physics, The University of Sydney, Sydney 2006, Australia}
\affiliation{$^{\dagger}$These authors contributed equally to this work}

\date{\today}
\begin{abstract}


A goal of the emerging field of quantum control is to develop methods for
quantum technologies to function robustly in the presence of noise. Central
issues are the fundamental limitations on the available information about
quantum systems and the disturbance they suffer in the process of
measurement. In the context of a simple quantum control scenario---the
stabilization of non-orthogonal states of a qubit against dephasing---we
experimentally explore the use of weak measurements in feedback control.  We
find that, despite the intrinsic difficultly of implementing them, weak
measurements allow us to control the qubit better in practice than is even
theoretically possible without them.  Our work shows that these more general
quantum measurements can play an important role for feedback control of
quantum systems.

\end{abstract}




\maketitle

Quantum technologies, such as quantum computing or quantum key distribution, offer many advantages over their classical counterparts, but it is a major challenge to make these new technologies function robustly in the presence of noise.  The field of \emph{quantum control}---the application of control theory to quantum systems---offers a spectrum of techniques to develop robust quantum technologies~\cite{WisemanMilburnBook, NJPcontrol}.  Control strategies can be efficient for driving quantum systems to a known target state, with applications ranging from phase estimation\cite{PhysRevLett.89.133602, higgins} and state discrimination~\cite{CookNature,Higgins2009}, to controlled state evolution in cavity QED \cite{PhysRevLett.89.133601}, to the cooling of single atoms \cite{bushev:043003} and macroscopic oscillators \cite{lahaye, gigan, heid, bouw, corbitt:160801}.

One such technique from control theory---feedback control---monitors the system and feeds back corrections onto it.  In classical feedback control strategies, it is always beneficial to acquire as much information about the system as possible, in order to identify the best correction.  However, this approach is not generally appropriate for the control of quantum systems, where two features of quantum measurement become important.  First, Heisenberg's uncertainty principle imposes fundamental limits on the amount of information that can be obtained about a quantum system, even in principle. Second, the act of measurement necessarily disturbs the quantum system in an unpredictable way. These fundamental features of quantum mechanics require a reevaluation of conventional methods and techniques from control theory when developing the theory of quantum control.  In particular, it suggests the use of variable strength or `weak' measurements~\cite{PhysRevA.41.11}, which balance the trade-off between information gain and disturbance in quantum systems.

In this Letter, we experimentally investigate a control scheme of the smallest possible quantum system---a qubit---for which weak measurements are necessary to achieve optimal control.  Motivated by the proposal of Branczyk \textit{et al.}~\cite{branczyk:012329}, we demonstrate the stabilization of a single qubit, prepared in one of two non-orthogonal states, against dephasing noise~\cite{NC01}.  Using weak measurements and active feedback, we achieve an improvement over even the theoretical best performance when using either strong projective measurements, or schemes without any measurement.  In contrast to control schemes conditioned on one outcome of a weak measurement (and failure for the other outcome)~\cite{katz:200401}, our scheme uses active feedback and succeeds for any weak measurement outcome.

We choose two non-orthogonal states because they serve as the simplest set of inputs demonstrating the limitation imposed by quantum measurement:  due to their non-orthogonality, it is impossible to design a control procedure that can perfectly discriminate the input state~\cite{WisemanMilburnBook} and subsequently control the resulting (known) input against noise. Non-orthogonal qubit states are particularly important in the well-known B92 quantum key distribution protocol~\cite{B92}.

\begin{figure}[ht]
\includegraphics[width=1 \columnwidth]{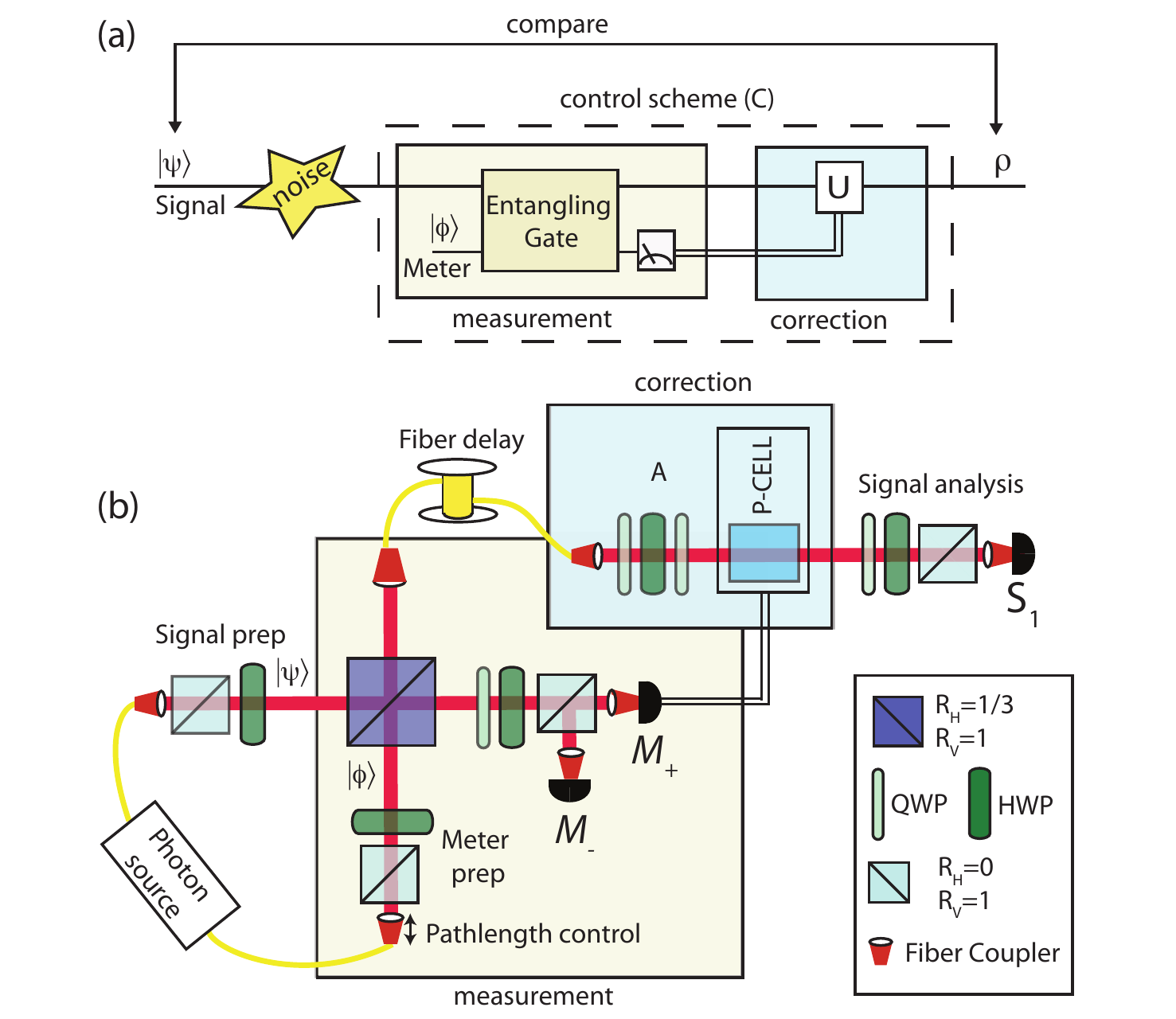}
\caption{\label{circuit}
(a) Schematic of a generic quantum feedback control procedure. A `signal' qubit passes through a noisy channel, is subsequently measured and corrected based on the result. Our scheme is shown schematically in the boxes.
(b) Experimental diagram. Qubits are encoded in the polarisation of single photons, (Horizontal $\ket{H}{\equiv}\ket{1}$, Vertical $\ket{V}{\equiv}\ket{0}$)  and manipulated with half waveplates (HWP) and quarter waveplates (QWP). Optical modes are overlapped on a partially polarising beam splitter. with reflectivities $R_H=1/3$ ($R_V=1$) for horizontally (vertically) polarised light . Conditional on there being only 1 photon in each input and output mode, the gate ideally applies the operation $\ket{0}\bra{0}{\otimes}{\mathbbm{1}}{+}\ket{1}\bra{1}{\otimes}Z $~\cite{langford:210504, kiesel:210505, okamoto:210506}. The outcome of a projective measurement of the meter in the +/- basis~\cite{note1}, using a polarising beamsplitter ($R_H=0$, $R_V=1$), determines a correction rotation on the signal qubit. Wave plate set A corrects for unwanted polarisation rotations induced by the fiber delay.  A coincident detection event between either ($M_+$ \& $S_1$) or ($M_-$ \& $S_1$) signals a successful run. Photon pairs are generated via spontaneous parametric downconversion ($\textsc{spdc}$) of a type I BiBO crystal cut to produce degenerate 820nm photon pairs. The crystal is pumped by a ${\approx}100$mW beam at 410nm produced from second harmonic generation of an 820nm mode-locked Ti:Sa laser (rep. rate 80MHz). We spectrally filter the \textsc{spdc} using $\pm{1}$nm interference filters centered at 820nm, then couple into two single mode fibres. Pump power is kept low so to reduce the presence of multi-pair emission from \textsc{spdc}~\cite{Tills, JMOpower}. \textsc{p-cell} is a Pockels cell.}
\end{figure}

Consider the stabilization of a qubit, which we will call the `signal', prepared in one of two non-orthogonal states $\ket{\psi_{\pm}}{=}\cos\frac{\theta}{2}\ket{+}{\pm}\sin\frac{\theta}{2}\ket{-}$~\cite{note1}, where $2\theta$ represents the angle between these two states on the Bloch sphere.  The signal qubit is transmitted through a noisy quantum channel that causes dephasing~\cite{NC01}, i.e., with probability $p$ a phase-flip operator $Z$ (where $Z\ket{0}{=}\ket{0}, Z\ket{1}{=}{-}\ket{1}$) is applied to the system (where $0{\le}p {\le}1/2$). This process can be described by a quantum operation $\rho_\pm {\to} \rho'_\pm$ on the initial state  $\rho_\pm{=}\ketbra{\psi_\pm}{\psi_\pm}$, given by
\begin{equation}
\label{noise}
  \rho^{'}_{\pm}= (1-p) \rho_{\pm}+p Z \rho_{\pm} Z.
\end{equation}

We seek a control procedure, described by a map $\mathcal{C}$, that can return the signal as close as possible to its initial noiseless state, independent of which initial state was prepared; see Fig.~\ref{circuit}a.  To quantify the performance of $\mathcal{C}$ we use the average fidelity $\overline{F}$ between the noiseless input state and the corrected output state.

\begin{align}
\overline{F}&=\tfrac{1}{2}[\bra{\psi_+}\mathcal{C}(\rho'_+)\ket{\psi_+}+\bra{\psi_-}\mathcal{C}(\rho'_-)\ket{\psi_-}]\\
&=\tfrac{1}{2}(F_{\psi_+}+F_{\psi_-})\,.
\end{align}

A general quantum feedback control scheme allows for a variable-strength, non-destructive, measurement of the qubit state. The result of this measurement then determines the sense of a correction rotation.  We now develop an alternate scheme \cite{robin} to the one presented in Ref.~\cite{branczyk:012329}.  In particular, we choose a family of weak measurement that, in addition to allowing for optimal control, exhibits interesting limiting cases with which to compare performance.  Consider a measurement in the logical basis $\{|0\rangle,|1\rangle\}$, with measurement operators~\cite{NC01}
\begin{align}\label{eq:M+}
  M_+ &= \cos(\chi/2)|0\rangle\langle 0| +  \sin(\chi/2)|1\rangle\langle 1|\,, \\
 \label{eq:M-} M_- &= \sin(\chi/2)|0\rangle\langle 0| + \cos(\chi/2)|1\rangle\langle 1|\,,
\end{align}
where $\chi$ ranges from 0 to $\pi/2$. The corresponding positive operator-valued measurement operators~\cite{NC01} are given by $\Pi_\pm{=}M_\pm^\dag M_\pm {=}[\mathbbm{1}\pm \cos{(\chi)}Z]/2$, where $\mathbbm{1}$ is the identity operator. 
 Defining the measurement strength as $\cos{\chi}$, we have $\chi{=}\pi/2~(\chi{=}0)$  corresponding to no measurement (projective measurement). Based on the weak measurement outcome $\pm$, we then perform a correction rotation $Y_{\pm\eta}$ by an angle $\pm\eta$ around the $y$-axis of the Bloch sphere, i.e., $Y_{\pm\eta} = \exp(\pm i\eta Y)$~\cite{note5}.

The average fidelity of this scheme can be optimized (analytically) over the correction angle, yielding

\begin{multline}
\label{poop}
\overline{F}(\theta,p,\chi)= \tfrac{1}{2} + \tfrac{1}{2}\Bigl[\bigl\{1{-}(1{-}(1{-}2p)\sin\chi)\cos^2\theta\bigr\}^2 \\ +\cos^2\chi\cos^2\theta\Bigr]^{1/2}\,,
\end{multline}
for a correction angle
\begin{equation}
\label{cheese}
\eta_{\rm opt}(\theta, p, \chi)= \tan^{-1}  \frac{\cos\chi\cos\theta}{1{-}(1{-}(1{-}2p)\sin\chi)\cos^2\theta} \,.
\end{equation}

\noindent In the case of zero strength ($\chi{=}\pi/2$), no measurement is performed, and Eqn.~\eqref{cheese} yields an optimal correction rotation of $\eta{=}0$. We call this the `do-nothing'  (DN) control scheme. In the case of maximum strength projective measurement ($\chi{=}0$), there is in general a non-zero optimal correction rotation angle.  This measurement in this scheme is the well-known Helstrom measurement~\cite{helstrom} that achieves the highest probability of discriminating between the two non-orthogonal states $\rho'_{\pm}$, and so we refer to this as the `Helstrom' (H) scheme.

\begin{figure}[t]
\includegraphics[width=1 \columnwidth]{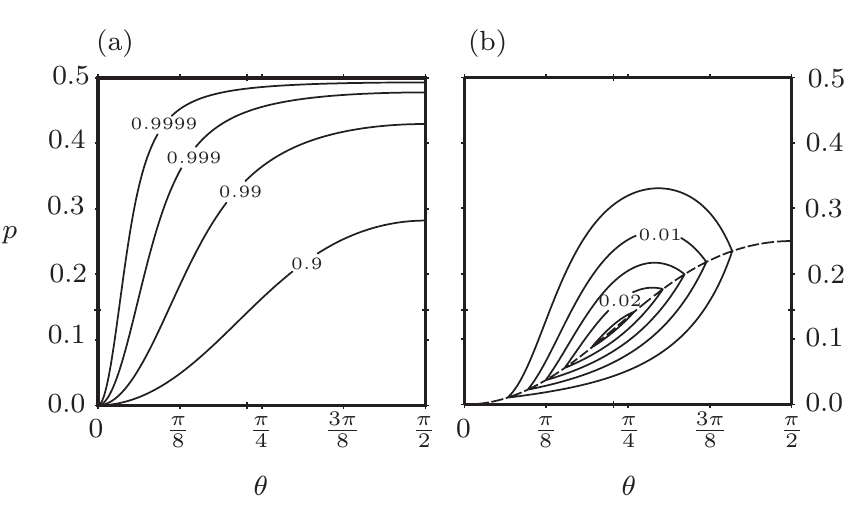}
\vspace{-6mm}
\caption{\label{3D}
(a) Contour plot of the optimal measurement strength $\cos{\chi_{\rm opt}}$ as a function of $p$ (the noise probability) and $\theta$ (half the angle between the two non-orthogonal qubit states, on the Bloch sphere).
(b) Contour plot of $\overline{F}_{\rm diff}$ representing the improvement to the control scheme achieved by allowing for variable strength measurements, see text.
The dashed line shows the cases where $\overline{F}_{DN}{=}\overline{F}_{H}$. Below this line $\overline{F}_{DN}{>}\overline{F}_{H}$ i.e. the Helstrom scheme (H) performs \emph{worse} than doing nothing at all. This clearly emphasises that strong measurements are not generally appropriate for the control of quantum systems.}
\label{theory}
\vspace{-2mm}
\end{figure}

In general, neither the DN scheme nor the H scheme correspond to the optimal control protocol.  Maximizing Eqn.~\eqref{poop} with respect to $\chi$ yields
\begin{equation}
\label{chiopt}
\chi_{\rm opt}(\theta,p) =\sin^{-1}\frac{(1-2p)\sin^2\theta}{1-(1-2p)^2 \cos^2\theta}\,.
\end{equation}
Fig.~\ref{theory}a shows a contour plot of the optimal measurement strength $\cos{\chi_{opt}}$.  Except for limiting cases of the initial conditions ($\theta{=}\{0, \pi/2\}, p{=}\{0,0.5\}$) neither the H nor DN schemes are optimal, i.e., in general \emph{there is always a non-trivial measurement strength which optimises control performance}. Substituting Eqn.~\eqref{chiopt} into Eqn.~\eqref{poop} yields a final expression for the optimum fidelity:
\begin{equation}
\label{poop2}
\overline{F}_{\rm opt}(\theta,p)= \tfrac{1}{2} + \tfrac{1}{2}\Bigl[ \cos^2\theta + \frac{\sin^4\theta}{1{-}(1{-}2p)^2\cos^2\theta}\Bigr]^{1/2}\,.
\end{equation}
In Ref.~\cite{branczyk:012329}, it was shown that this average fidelity is the \emph{optimal} performance that can be achieved by \emph{any} feedback control scheme (all possible physical maps).

To quantify the improvement offered by variable strength measurements we consider the quantity $\overline{F}_{\rm diff}{=}\overline{F}_{\rm opt}{-}\text{max}\{\overline{F}_{H},\overline{F}_{DN}\}$, which is graphed in Fig~\ref{theory}b. The largest improvement,  $F_{\rm diff}{=}0.026$, occurs at $\theta{=}0.715$ and $p{=}0.115$.

We implement the general control strategy in a photonic architecture, encoding qubits into the polarisation of single photons, see Fig.~\ref{circuit}b. The required variable strength quantum non-destructive measurement  on the signal qubit is realised by entangling it to another `meter' qubit (photon) using a non-deterministic linear optic controlled-\textsc{z} (\text{cz}) gate~\cite{langford:210504, kiesel:210505, okamoto:210506}, then performing a full strength projective measurement on the meter. This implements a measurement on the signal (Eqns.~\eqref{eq:M+}-\eqref{eq:M-}) with a strength determined by the input meter state $\ket{\phi}{=}\cos\frac{\chi}{2}\ket{+}{+}\sin\frac{\chi}{2}\ket{-}$, as described in~\cite{PhysRevLett.92.190402}.

The sign of the correction rotation on the signal photon, $\pm \eta$ of Eqn.~\eqref{chiopt}, is determined by the outcome of the meter measurement, i.e., whether detector $M_+$ or $M_-$ fires (see Fig.~\ref{circuit}b). We implement this correction using a 3mm Rubidium Titanyl Phosphate (RTP) Pockels cell with a half-wave voltage of $\sim$700 V at 820 nm.  A similar   scheme to perform classical feed-forward based on the outcome of a measurement was implemented in \cite{sciarrino} which allowed for a minimum-disturbance measurement of polarisation encoded qubits. We adjust the correction $\eta$ by varying the applied  voltage. Rather than requiring the Pockels cell to implement two different rotations ($\pm\eta$) we use fixed wave plates (labelled signal analysis in Fig.~\ref{circuit}b) to always rotate the signal by $-\eta$, and the Pockels cell to rotate by $+2\eta$ conditional on the meter outcome $M_{+}$. To allow time for the outcome of the meter measurement to be processed and to trigger the Pockels cell, we send the signal photon through 50m of optical fiber. The birefringence of this fiber causes unwanted polarisation rotations which are removed using additional wave plates, see Fig.~\ref{circuit}b.

Noise on the signal state is implemented by making use of the decomposition of Eqn.~\eqref{noise} into an ensemble of pure states $\{\rho_\pm\,, \, Z\rho_{\pm}Z\}$, weighted by the noise probability.  For each measurement strength, we perform state tomography on the signal output for each input state $\rho_\pm$ and $Z\rho_\pm Z$ separately. The count rates are then combined, weighted by the noise probability $p$, and the noisy density matrix is reconstructed via a linear inversion~\cite{PhysRevA.64.052312,LangfordNK2007phd}. At the output of the optical circuit we observe a count rate of approximately 100 coincident photon pairs per second.

\begin{figure*}[ht]
\includegraphics{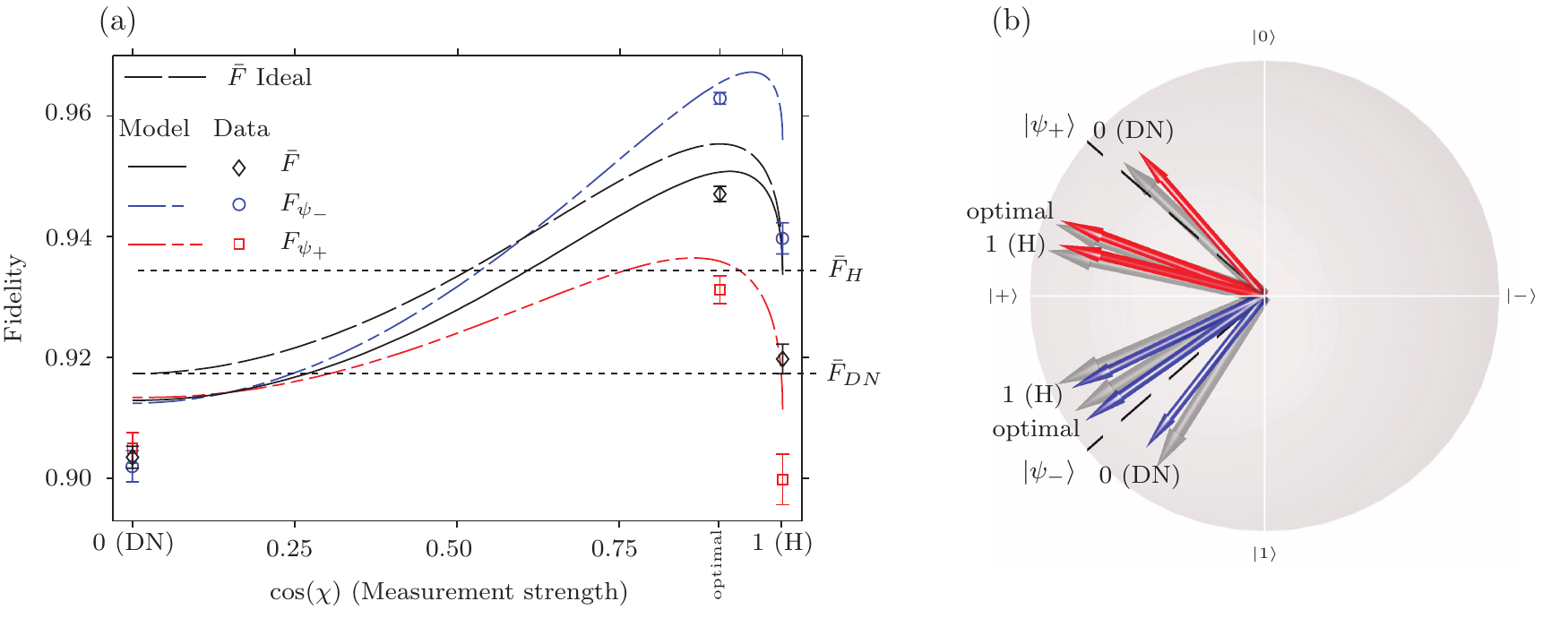}
\vspace{-4mm}
\caption{\label{results}
Experimental results of our quantum feedback control procedure. 
(a) Fidelities between input and output states (Eqns 2-3), as a function of measurement strength ($\cos{\chi}$). The model is discussed in the text.  Error bars are calculated from poissonian photon counting statistics.
(b) Corresponding output states presented as vectors on the Bloch sphere. Light grey vectors are those predicted by the experimental model, as in panel a. of the figure.
The coloured vectors correspond to coloured points in panel a., and the corresponding measurement strength is shown next to each vector.
}
\vspace{-2mm}
\end{figure*}

We implement the feedback scheme at $\theta{=}0.715$, $p{=}0.145$ (a point at which our theoretical analysis predicts close to the largest improvement $\overline{F}_{\rm diff}$, Fig.~\ref{theory}b) and three measurement strengths: zero ($\cos{\chi}{=}0$, the DN scheme); the theoretical optimal value ($\cos{\chi_{opt}}$); and maximum ($\cos{\chi}{=}1$, the H scheme). Fig~\ref{results}a compares the experimental results with the ideal and a model that includes experimental imperfections. This model allows for measured imperfections in the reflectivities of the central  beamsplitter in the entangling (\textsc{cz}) gate~\cite{langford:210504, kiesel:210505, okamoto:210506}, Fig.~\ref{circuit}b, which deviated from ideal by around a percent ($R_H{=}0.345, R_V{=}0.995$). 
The close fit between our experimental model and results shows that the main limiting factors in our experiment are imperfections in the properties of manufactured optical elements. We attribute the remaining differences between the model and data to thermal drift during experiments. 

The key result is that our experimental control protocol performed best (on average and for each individual state) when employing an intermediate strength measurement: the highest measured average fidelity achieved is $\bar{F}_{opt}{=}0.947{\pm}{0.001}$ at a measurement strength of $\cos{\chi}{=}0.93$. Furthermore, this result is higher than even the \emph{theoretical} best performance of the limiting schemes ($F_H{=}0.9344{>}F_{DN}$), i.e., despite experimental imperfections, the use of intermediate strength measurements produce higher fidelity than a \emph{perfect} experiment constrained to using the DN or H strategies. Fig.~\ref{results}b presents the six states, from which the fidelities in Fig.~\ref{results}a were deduced, as vectors on the Bloch sphere.

In conclusion, we have demonstrated quantum feedback control on a single quantum system - a photonic polarization qubit. We demonstrated the use of weak measurement in obtaining the optimal tradeoff between information gain and measurement back-action. This motivates the investigation of generalized measurements for control protocols in a range of quantum systems, where quantum control will ultimately be used for realizing quantum technologies.

We acknowledge Agata Bra\'{n}czyk, Andrew Doherty and Alexei Gilchrist for valuable discussions. This research was supported by the Australian Research Council.
\vspace{-5mm}

\begin{thebibliography}{28}
\bibitem[{not()}]{note1}
\bibinfo{howpublished}{We have defined $\ket{\pm}=\frac{1}{\sqrt{2}}\left(
  \ket{0}\pm\ket{1} \right)$, where $\ket{0}$ and $\ket{1}$ represent the
  computational basis of the qubit.} 
\bibitem[{not()}]{note5}
\bibinfo{howpublished}{With our choice of noise and subsequent weak measurement, the Bloch vector is at all times in the $x{-}z$ plane.  A $Y$-rotation is therefore a natural candidate for the correction, and we subsequently show that this choice leads to optimal performance. } 
\end{thebibliography}

\end{document}